\documentclass[
preprint,
showpacs,preprintnumbers,
 amsmath,amssymb,
 aps,
]{revtex4-1}
\usepackage{graphicx}
\usepackage{dcolumn}
\usepackage{bm}

\begin{document}

\preprint{APS/123-QED}

\title{Electron acceleration in a nonrelativistic shock with very high Alfv\'en Mach number }

\author{Y. Matsumoto}
\email{ymatumot@astro.s.chiba-u.ac.jp}
\affiliation{%
Department of Physics, Chiba University, 1-33 Yayoi-cho, Inage-ku, Chiba 263-8522, Japan}%


\author{T. Amano}
\author{M. Hoshino}
\affiliation{
 Department of Earth and Planetary Science, the University of Tokyo, 7-3-1 Hongo, Bunkyo-ku, Tokyo 113-0033, Japan}%


\date{\today}

\begin{abstract}
Electron acceleration associated with various plasma kinetic instabilities in a nonrelativistic, very-high-Alfv\'en Mach-number ($M_A \sim 45$) shock is revealed by means of a two-dimensional fully kinetic PIC simulation. Electromagnetic (ion Weibel) and electrostatic (ion-acoustic and Buneman) instabilities are strongly activated at the same time in different regions of the two-dimensional shock structure. Relativistic electrons are quickly produced predominantly by the shock surfing mechanism with the Buneman instability at the leading edge of the foot. The energy spectrum has a high-energy tail exceeding the upstream ion kinetic energy accompanying the main thermal population. This gives a favorable condition for the ion acoustic instability at the shock front, which in turn results in additional energization. The large-amplitude ion Weibel instability generates current sheets in the foot, implying another dissipation mechanism via magnetic reconnection in a three-dimensional shock structure in the very-high-$M_A$ regime.
\end{abstract}

\pacs{52.35.Tc, 52.65.Rr, 96.50.Pw, 98.70.Sa}
\maketitle

Collisionless shocks provide us great opportunities to explore nonlinear dynamics in strongly inhomogeneous plasmas. Dynamics therein result in excitation of various types of electrostatic and electromagnetic waves and associated plasma heating and acceleration. Extreme circumstances encountered in such situations can be realized in astrophysical phenomena, such as supernova remnant (SNR) shocks where the plasma kinetic energy overwhelms other magnetic and plasma internal energies. SNR shocks have been thought to be a generator of cosmic rays, and exploring nonlinear dynamics in extreme circumstances therefore clarifies how charged particles are accelerated to relativistic energies out of the thermal counterpart.

The magnetized collisionless shock is characterized by the Alfv\'en Mach number $M_A$, which is the ratio of the flow speed $V_0$ to Alfv\'en speed $V_A$ in the upstream. When the Alfv\'en Mach number exceeds a critical value ($\sim 3$), the plasma cannot be fully dissipated at the shock, and additional dissipation is compensated by the ion specularly reflected by the shock front \cite{Tidman_Krall_1971,Leroy_1983}. In particular, in very-high-$M_A$ shocks, the ion can provide free energy for various plasma kinetic instabilities \cite{Wu_1984,Papadopoulos_1988,Cargill_Papadopoulos_1988}. 

SNR shocks are indeed such cases  of very high $M_A$. Remote imaging of SNR shocks has provided rich information of fine-scale structures in which the presence of relativistic electrons has been evidenced \cite{Reynolds_2008}. It is only recently that a high-$M_A$ shock was directly measured with relativistic electrons accelerated in the vicinity of the Kronian bow shock \cite{Masters_2013}. Laboratory experiments involving a high-power laser facility provide other opportunities of exploring high-Mach-number shocks \cite{Silva_2004,Park_2012,Kuramitsu_2012}. Although such experimental studies reveal the macroscopic nature of high-Mach-number shocks, they still lack detailed information on electric and magnetic fields, and associated mechanisms of particle acceleration.

While numerical simulation is an alternative way of exploring these extreme environments, examining nonrelativistic, high-$M_A$ shocks is still computationally challenging. This is because of a strong dependence of CPU time on the ion-to-electron mass ratio ($M/m$), which increases as $(M/m)^{3 (2.5)}$ in a three- (two-)dimensional fully kinetic simulation of a collisionless shock. This scaling has limited discussions either with small mass ratios or in moderate $M_A$ shocks. Nonetheless, numerical experiments have revealed part of their signatures. One-dimensional simulation studies have proposed an efficient electron acceleration mechanism in the high-$M_A$ regime, which can be an agent of the diffusive shock acceleration in SNR shocks \cite{MacClements_2001,Hoshino_Shimada_2002}. The process - the electron shock surfing acceleration (SSA) - is realized with an electrostatic field via the Buneman instability and the motional electric field at the leading edge of the foot. The accelerated electrons are trapped/reflected by large-amplitude wave electric fields, in contrast to the classical shock drift acceleration where the compressed magnetic field plays the role 　\cite{Krauss-Varban_1989}, and are energized much more efficiently. It has been a controversial issue whether the mechanism operates efficiently in multidimensional shock structures \cite{Ohira_Takahara_2007,Amano_Hoshino_2009a,Riquelme_Spitkovsky_2011}. Investigation of nonlinear saturation levels of the Buneman instability in multiple dimensions \cite{Amano_Hoshino_2009b} led to the condition 
\begin{equation}
M_A \gtrsim (M/m)^{2/3}
\label{ma_eq}
\end{equation}
for the effective electron SSA \cite{Matsumoto_2012}. There is therefore a need for very-high-$M_A$ shock studies with more physically important mass ratios. In this Letter, we report results from a fully kinetic particle-in-cell (PIC) simulation of such a very-high-$M_A$ shock. We found that various types of electrostatic and electromagnetic instabilities are strongly activated at the same time, but in different regions of the two-dimensional shock structure. Electron energization associated with the instabilities is discussed.

\begin{figure*}
\includegraphics[scale=0.5]{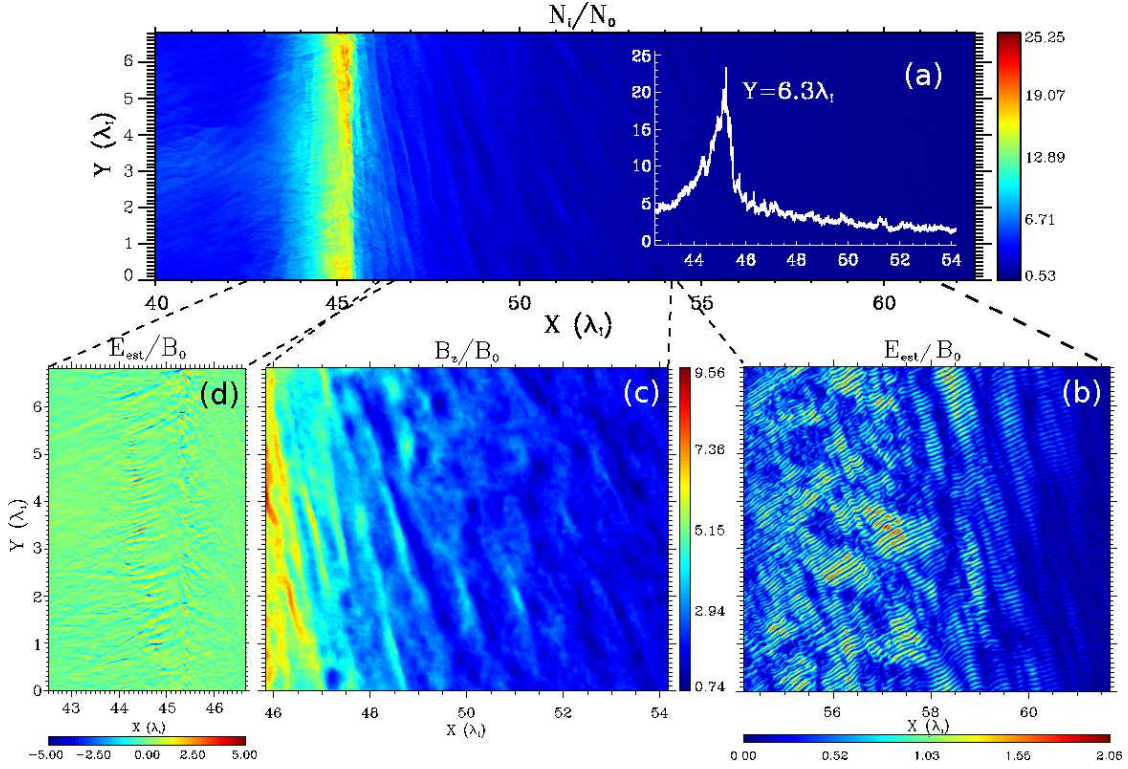}
\caption{Overview of the two-dimensional shock structure at $\Omega_{gi}T=4$. (a) Two-dimensional profile of the ion number density normalized by the upstream value. The inset shows the profile along the x axis at $Y=6.3 \lambda_i$. (b) Strength of the electrostatic field at the leading edge of the foot in the zoomed-in region in panel (a). (c) $B_z$ in the foot. (d) The y component of the electrostatic field around the overshoot. The fields are normalized by the upstream magnetic field $B_0$.}
\label{fig:ov}
\end{figure*}

We use a two-dimensional PIC code to examine a shock evolution. The code implements a second-order (spline) shape function with a charge conservation scheme \cite{Esirkepov_2001} to inhibit ``low-frequency'' numerical Cherenkov radiation in the upstream caused by aliasing errors of the shape function. The code is hybrid parallelized by one-dimensional domain decomposition using a message passing interface (MPI) library and OpenMP, and optimized for recently developed massively parallel supercomputer systems. To create a collisionless shock, we adopt the injection method in which particles are continuously injected from one boundary ($x=L_x$) with a speed $-V_0$ toward the other end ($x=0$) where the particles are reflected, resulting in a shock formation that propagates toward $x=L_x$. The number density in the upstream $N_0$ is 40 particles per cell for each species (ion and electron). The injected plasma carries the z component of the magnetic field ($B_0$) and the motional electric field $E_y = E_0=-V_0 B_0/c$. Thus, we deal with a purely perpendicular shock in the downstream rest frame. The periodic boundary condition is applied in the y direction. The simulation box size in the x direction is expanded as the shock wave propagates in time. The size in the y direction is $L_y=6.8\ \lambda_i$, where $\lambda_i=c/\omega_{pi}$ is the ion inertia length in the upstream. The grid size $\Delta h$ and the time step size $\Delta t$ are set as $\Delta h = \lambda_D$ and $\omega_{pe} \Delta t=0.025$, where $\lambda_D$ is the Debye length and $\omega_{pe}$ is the electron plasma frequency in the upstream. We carried out a simulation run with a mass ratio $M/m=225$ and Alfv\'en Mach number in the shock rest frame $M_A = 44.8$, satisfying Eq. (\ref{ma_eq}). The electron plasma $\beta_e=0.5$, the ratio of the electron plasma to gyro frequencies $\omega_{pe}/\Omega_{ge}=10.0$, and the temperature ratio of the ion to the electron $T_i/T_e=1$. In the late stage of the run, $\sim10^{10}$ particles are followed in the simulation domain with $24000 \times 2048$ grid points.

The spatial profile of the ion number density at $\Omega_{gi}T = 4$ is shown in Fig. \ref{fig:ov}(a). From the upstream (right) to the downstream (left) regions, there is a transition between $X=60 \lambda_i$ and $X=47 \lambda_i$ (foot). This transition region is followed by a rapid increase in the number density (ramp), a peak of the value (overshoot), and a recovery to the downstream value ($\sim 4N_0$) in $X < 42 \lambda_i$. While this overall signature is essentially the same as signatures found in supercritical collisionless shocks \cite{Leroy_1983}, the magnitude of the overshoot value reaches $N_i=25N_0$. A long-wavelength mode (m = 1) as seen in the density inhomogeneity along the shock front has been similarly found in the two-dimensional kinetic hybrid simulations \cite{Burgess_Scholer_2007}.

\begin{figure*}
\includegraphics[scale=0.5]{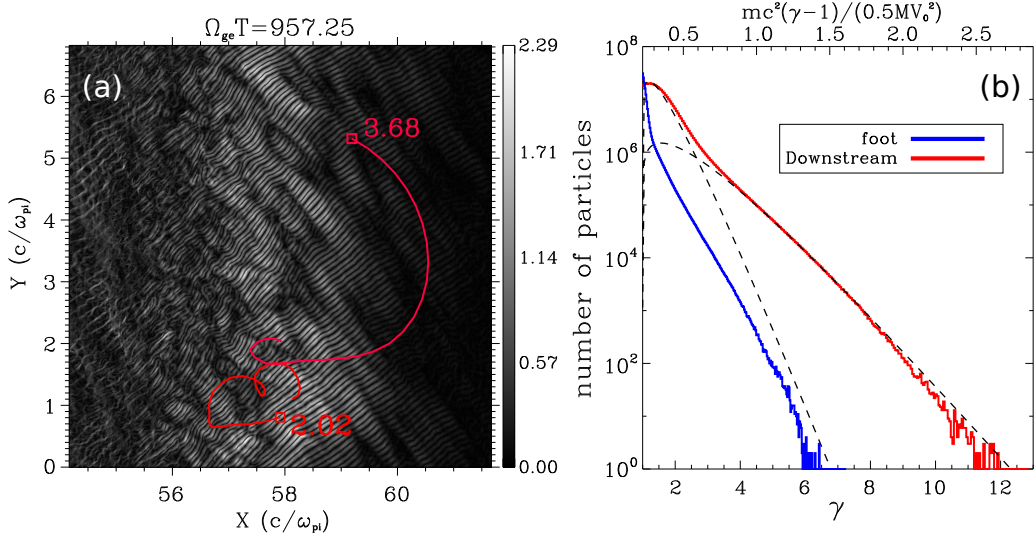}
\caption{(a) Typical trajectories of accelerated electrons (red) superposed on the two-dimensional profile of the electrostatic field strength normalized by $B_0$. Open squares indicate the positions of the particle following the trajectory back for $15\ \Omega_{ge}^{-1}$. The number next to each open square is $\gamma$. The snapshot is taken at $\Omega_{ge}T=957.25$. (b) Energy spectra of the electron in the foot (blue) and downstream (red) regions. Dashed lines are fitted relativistic Maxwell distributions to the downstream spectrum. The lower and upper x axes refer to $\gamma$ and the kinetic energy normalized by the upstream ion kinetic energy, respectively.}
\label{fig:accel}
\end{figure*}

Structures of different scales and modes are found in these regions. Figure \ref{fig:ov}(b) shows the strength of the electrostatic field $|{\bf E}_{est}| = |-{\bf \nabla} \phi|$, where $\nabla^2 \phi = -\nabla \cdot \bf E$, at the leading edge of the foot. The waveform is coherently aligned with a wavelength much smaller than the ion inertia length, the electron scale. The wave vector is directed oblique to the x and y axes. This electrostatic wave is a result of the Buneman instability excited by interaction between the upstream electron and the reflected ion.

 Some electrons are quickly accelerated when they enter the foot by the strong electrostatic field. Figure \ref{fig:accel}(a) shows typical accelerated electron orbits plotted over the electrostatic field strength. The amplitude of the wave at the leading edge is twice as large as the local magnetic field strength ($\sim B_0$), satisfying the condition of the unlimited SSA \cite{Hoshino_Shimada_2002,Shapiro_Ucer_2003}. The potential well captures the electrons as they are accelerated by the motional electric field in the y direction. The electron is accelerated to relativistic energies with a Lorentz factor of $\gamma \sim 4$ within the time scale of the electron gyro period. A nonthermal electron energy distribution has already formed in the foot ($50.0 \lambda_i \le X \le 53.3 \lambda_i$) before reaching the shock front (blue line in Fig. \ref{fig:accel}(b)). In the downstream ($43.3 \lambda_i \le X \le 45.0 \lambda_i$, red line in Fig. \ref{fig:accel}(b)), the maximum energy reached $\gamma \sim 12$, which is 2.5 times the upstream ion kinetic energy in this particular case. Further acceleration after the electron SSA is basically adiabatic. The least-squres fitting by a double relativistic Maxwell distribution (dashed lines) gives temperatures of $T_1/mc^2=0.27$ for the main component and $T_2/mc^2=0.58$ for the nonthermal component. The energy distribution does not change much as time proceeds in $\Omega_{gi}T > 4$.

\begin{figure}
\includegraphics[scale=0.3]{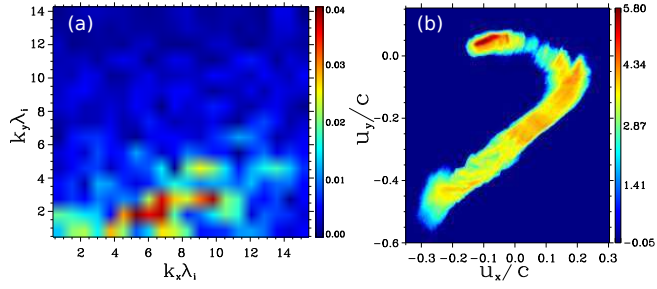}
\caption{(a) Fourier power spectrum of $B_z$ in the foot. The wave number on each axis ($k_x$ and $k_y$) is normalized by the ion inertia length. (b) Ion velocity distribution sampled in $46.5 \lambda_i \le X \le 48.5 \lambda_i$ and $2.4 \lambda_i \le Y \le 4.4 \lambda_i$. The axes are normalized by the speed of light. The number of particles in each bin is color coded on a logarithmic scale.}
\label{fig:iwbl}
\end{figure}

Figure \ref{fig:ov}(c) shows strong magnetic field perturbations in the entire region of the foot. The wave vector is almost orthogonal to the electrostatic mode at the leading edge of the foot in Fig. \ref{fig:ov}(b). The amplitude is very large as compared with the upstream magnetic field $B_0$. It varies from $5\ B_0$ to $10\ B_0$ within the ion inertia scale around $X=47 \lambda_i$, resulting in self-generated current sheets. The Fourier power spectrum of $B_z$ in the foot region in Fig. \ref{fig:iwbl}(a) shows that the mode with the wavelength of the ion inertia length ($|k|\lambda_i \sim 2\pi$) is dominant and the wave vector is tilted from the x axis, which are features closely related to the motion of the reflected ion. Figure \ref{fig:iwbl}(b) shows the ion distribution function sampled in $46.5 \lambda_i \le X \le 48.5 \lambda_i$ and $2.4 \lambda_i \le Y \le 4.4 \lambda_i$. There are two components. The relatively cold clump flowing in the -x direction is the incident ion. This component is slightly heated as it enters the destabilized region. The other component is the reflected ion exhibiting a gyrating motion in the velocity space. Thus, the velocity distribution function is highly anisotropic. This situation is subject to the ion-beam Weibel instability \cite{Kato_Takabe_2010b}. Indeed, the wave vector in Fig. \ref{fig:ov}(c) and Fig. \ref{fig:iwbl}(a) is almost perpendicular to the direction of the anisotropy in the velocity space. The observed ion-scale electromagnetic mode corresponds to the fastest growing mode of the instability \cite{Kato_Takabe_2010}. 

\begin{figure*}
\includegraphics[scale=0.5]{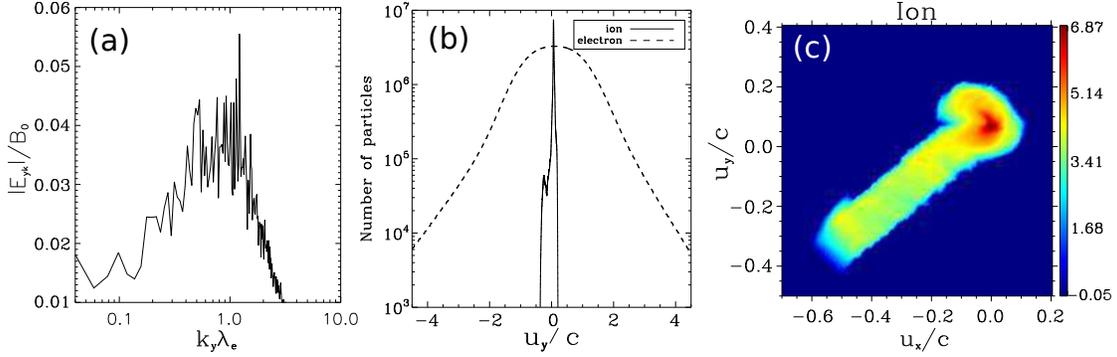}
\caption{(a) Fourier power spectrum of $E_y$ averaged in the downstream region ($43.3 \lambda_i \le X \le 45.0 \lambda_i$). The wave number $k_y$ is normalized by the average electron inertia length. (b) Distribution functions of the ion (solid line) and electron (dashed line) sampled in the downstream region. The x axis shows the y component of the four-velocity normalized by the speed of light. (c) Ion velocity distribution in the comoving frame of the shock front in the same format as Fig. \ref{fig:iwbl}(b).}
\label{fig:undr}
\end{figure*}

There exist large density gradients at the ramp and behind the overshoot in the present high-$M_A$ shock. The strong plasma inhomogeneity permits a kind of drift wave to grow along the shock surface with an amplitude of $|E| \sim 5B_0\  (25E_0)$ as shown in Fig. \ref{fig:ov}(d). Figure \ref{fig:undr}(a) shows the power spectrum of $E_y$ in the y direction averaged over the region behind the overshoot ($43.3 \lambda_i \le X \le 45.0 \lambda_i$). The excited strong electrostatic wave is powered at $k_y \lambda_e\sim1$, where $\lambda_e$ is the electron inertia length in this region. The region consists of the relativistically hot ($T_e \sim mc^2$) electron preheated by the Buneman instability at the leading edge of the foot (Fig. \ref{fig:undr}(b)), and the transmitted and reflected ions (Fig. \ref{fig:undr}(b) and \ref{fig:undr}(c)). The electron drift motion is in the +y direction with a speed of $v_d =0.1c$. Although the temperature ratio $T_e/T_i$ is not large, the non-Maxwell ion distribution and the background electron temperature gradient relax the threshold of the ion acoustic instability even for the case with $T_i \sim T_e$ \cite{Wu_1984,Priest_1972}.

The present configuration with the out-of-plane magnetic field component limits possibilities of other types of kinetic instability. In particular, the strong magnetic field compression at the overshoot would be free energy for the ion cyclotron instability owing to the anisotropy of the ion temperature \cite{Burgess_2006}. The resultant large-amplitude electromagnetic fields would modify the present coherent shock front structure and work as a scattering body for the preaccelerated electron \cite{Guo_Giacalone_2010}.

The unprecedentedly high-$M_A$ PIC simulation enabled us to confirm the theoretical prediction Eq. (\ref{ma_eq}) for the first time with a large mass ratio sufficient to separate ion and electron dynamics. Furthermore, the introduction of multidimensionality provides new insights into nonlinear shock dynamics, in which various kinetic instabilities are activated at the same time and competing with each other . The anisotropy of the ion distribution function in the foot destabilized the ion-beam Weibel instability that generates current sheets. This implies that the magnetic reconnection, which cannot be realized in the present two-dimensional configuration, can be another dissipation mechanism in shocks with much higher $M_A$ in three-dimensional space. The strongly inhomogeneous plasma around the overshoot introduces free energy for the drift instability along the shock surface. The characteristics suggest growth of the ion acoustic (IA) instability, while a number of instabilities have resulted from linear kinetic theories \cite{Lemons_1978,Wu_1984}. However, the instability works only for complementary heating of electrons, since they are already relativistically hot ($T \sim mc^2 \gg E_{\rm IA}^2/8\pi N_0$); electrons are substantially heated at the leading edge of the foot by the Buneman instability rather than at the shock front. The electrostatic field with large amplitude at the leading edge also allows efficient electron SSA. The resultant distribution of electron energy has a high-energy tail exceeding the upstream ion kinetic energy, suggesting that the electron SSA is a robust preacceleration mechanism that seeds the electron diffusive shock acceleration in young SNR shocks.

\acknowledgments
This work was supported by JSPS KAKENHI Grant-in-Aid for Young Scientists (Start-up) 23840047. Numerical computations were conducted using the Fujitsu PRIMEHPC FX10 System (Oakleaf-FX) at the Information Technology Center, The University of Tokyo.

%

\end{document}